# Sensitive detection of ultra-weak adhesion states of vesicles by interferometric microscopy

**Zen-Hong Huang[1], Gladys Massiera[2], Laurent Limozin[1], Paul Boullanger[3], Marie-Pierre Valignat[1], Annie Viallat[1]**

We used an original analysis of reflection interference contrast microscopy (RICM) to detect an ultra-weak specific interaction between a glycolipid vesicle and a lectin-coated substrate. The membrane height fluctuations in the contact zone are observed with high illumination aperture; the membrane profile and the membrane-substrate distance are quantitatively determined by using the new analysis, which accounts for multiple interfaces and multiple incidence rays. We showed that this refined version of RICM theory is necessary, specifically in the case of intermediate membrane-substrate distance (~30 nm) and helped to discriminate between ultra-weak interaction and pure gravitational sedimentation

## Introduction

Giant unilamellar lipid vesicles (GUVs) are widely used as biomimetic objects able to isolate and to modulate specific aspects of cell mechanics[1,2,3,4], dynamics[5,6,7,8,] or membrane organization[9,10]. As their membrane can contain several kinds of lipids bearing charges or specific functions, the vesicles can easily interact with a neighbour substrate, and this interaction may alter the vesicle behaviour. It is thus crucial to unambiguously detect and characterize possible membrane-wall interactions. Indeed, GUVs adhesion onto model substrates has been extensively studied[11,12,13,14,15], notably because it is a way to gain an insight into the physical basis of cell adhesion[16]. However, the case of very weak interactions is not very documented although in most physical and biophysical studies involving GUVs, the vesicles are located close to a substrate. Moreover, RICM theory and new experimental set ups such as the dual wave length method[17], while useful to extract

*1 Laboratoire Adhésion et Inflammation, U600-Inserm, UMR 6212, CNRS-Université Aix-Marseille 2, case 937, 163 Avenue de Luminy, 13288 Marseille, cedex 9, France Fax: 33 491 82 88 51 E-mail: annie.viallat@inserm.fr*
*2 Laboratoire des Colloïdes, Verres et Nanomatériaux, UMR 5587, CNRS/UM2, CC26, 34095 Montpellier Cedex 5, France*
*3 laboratoire Chimie Organique 2-Glycochimie, ICBMS –UMR CNRS 5246, Université Lyon 1 – Bat. Currien 43, Bd du 11 novembre 1918, 68622 Villeurbanne cedex France*

membrane profile for large membrane to substrate distances, have limited precision in the case of intermediate distances.

In this paper, we use for the first time a refined model of reflection interference contrast microscopy (RICM) to analyse RICM data obtained for vesicles in ultra-weak adhesion with a substrate[18]. This method allows to accurately reconstruct the vertical profile of the membrane close to the contact zone. We show that this method is necessary to enable the detection of ultra-weak adhesion states and to discriminate these states from that induced by pure gravitational settling. We stress that this method applies to vesicles which are not particularly deflated, for which the membrane thermal fluctuations may be of relatively weak amplitude.

RICM is the technique of reference for the observation of vesicle adhesion[17,19,20]. Based on the principle of Newton's rings formation, this method allows the visualization of the vesicle contact zone, and the detection of adhesion patches. Moreover, a simple theory, which relates intensity to height above the substrate, allows a quantitative analysis of the interference pattern and provides both the measurements of height fluctuations of the membrane and the reconstruction of the vertical profile of the membrane at the immediate vicinity of the substrate[15]. The technique has proved to be suitable to characterize membrane spreading times[13,22] membrane thermal fluctuations[21], membrane-substrate separation and associated adhesion strength[12,13,22].

However, in the context of the study of very weak adhesion, the theory used routinely for RICM analysis, which assumes two interfaces and normal incidence, is not valid, even when the membranes are roughly flat and parallel to the substrate. On the one hand, as outlined recently[22], the reflection on the internal side of the membrane induces an additional phase shift, which depends on the refractive indices of the internal vesicle medium. This shift sensitively changes the relation between membrane height and light intensity in the intermediate range of heights (30-50 nm) where weakly bound membranes are located. On the other hand, height fluctuations are most conveniently measured using a high illumination aperture in order to maximize the illumination. The interest is to reduce exposure time and blurring effects occurring with fast moving membrane parts. The consequence is that, even for relatively small membrane height (h~50nm), one observes a fringe contrast damping with the radial distance due to large angle incident rays[23]. The damping has to be accounted for to describe the variation of the fringes light intensity with distance and to obtain an accurate height reconstruction of the vesicle profile over several fringes.

We use in this study a theory of RICM, which accounts for multiple interfaces and multiple incidences rays. One key improvement is to account for interferences between rays of different incident polarization and also to carefully normalize the data. While the interfaces are supposed to be horizontal, the theory applies to the reconstruction of the vesicle profile up to 4-5 fringes, better than the current most elaborated theory[19]. We work on a statistical ensemble of glycolipid and phospholipid vesicles electroformed in standard conditions and used without further treatment (no significant deswelling stage for instance) in order to fit to many experimental situations encountered in vesicle studies. We show the existence of an ultra weak attraction between glycolipid vesicles and lectin surfaces, which we discriminate from the pure settling state presented by glycolipid vesicles on silanized surface and by phospholipid vesicles both on lectin and silanized surfaces.

**Background on vesicle specific adhesion**

Vesicle adhesion has been extensively studied both experimentally[24] and theoretically[25]. Usually, GUVs exhibit key elements of the cell surface involved in adhesion[26]. They are generally prepared with receptors or receptor fragments reconstituted into the membrane and with lipids bearing polymer headgroups of polyethylene-glycols[11,12,13,14,27] which mimic the glycocalix and model the associated generic long range repulsion force. The substrate is constituted by supported lipid bilayers containing homophilic receptors[12,28,29] or coated/grafted with ligands recognizing the receptors[13].

The specific adhesion of membranes to substrates is governed by an interplay of short range attractions between receptor-ligand pairs and non-specific repulsive interactions such as Helfrich entropic repulsion. As a result of this attraction/repulsion competition, three main behaviours have been reported[16]:

The non-adherent regime, determined by the balance of gravitational attraction (vesicle weight) and Helfrich repulsion. It is characterized by a strong membrane flickering in the adhesion disc. The contact area is inhomogeneous, and the probability distribution of the membrane-substrate distance in the contact disc, P(h), (estimated from RICM) is Gaussian with a large mean value h > 50nm, and a roughness (root mean square of P(h)) of about 25 nm[30].

The intermediate adherent regime, which occurs in presence of mobile receptors and repulsive lipopolymers. It is characterized by the formation of adhesion plaques, which are related to the lateral segregation of receptor-ligand pairs [29,31]. This regime results from the competition between short range ligand-receptor forces and long range glycocalix repulsion, which leads to spontaneous segregation of receptors. The tight adhesion plaques do not show fluctuations, and are associated with a membrane-substrate distance h < 30 nm. In contrast, the non-adherent regions of the contact zone are flickering. Recently, adhesion domains were detected in absence of repellers in the case of weak receptor-ligand interactions when the ligands are anchored to a polymeric spacer. Free ligands then provide a membrane-substrate repulsion comparable to that of the usual lipopymers (PEG2000) used as repellers [27].

The strongly adherent regime induced either by specific or non-specific forces. The contact area has an homogeneous quasi-circular non-fluctuating shape, characterized by a dark ring surrounding the adhesion disc[22] and a membrane-substrate distance[12] h < 10 nm.

In our system, the specific attraction caused by the recognition between a glycolipid and a lectin result in an ultra-weak adhesion state characterized by a small circular homogenous adhesion disk with an intermediate distance to the substrate.

**RICM model**

We have analysed RICM images by using a new and recently developed optical model detailed elsewhere[18]. This model quantitatively calculates the intensity reflected by the vesicle interface and by the substrate by taking into account three important specificities of the experimental system. The first one is the superposition of different refractive indices layers formed by the vesicle and the substrate (3 interfaces). The second one is the illumination numerical aperture (INA) of the microscope, responsible for a large angle distribution of incident rays. A summation of the intensity over all angles of illumination is therefore required to compute the light reflected intensity. The third characteristics is the polarization of the light. Indeed, the presence of the antiflex component in the experimental set up induces light polarization, which had never been accounted for so far, even in the most comprehensive models (Wiegand[19], Gingell[23]).

In the analysis presented here, the total reflected intensity is expressed as a function of the amplitude of the reflection coefficients of a multifilm structure, $R_s$ and $R_p$. It can be written as the sum of a first term, equal to the intensity obtained with unpolarized light and corresponds to the intensity calculated by Wiegand[19] for plane interfaces and of a second corrective term due to the polarization :

$$I = \int_0^{\theta_{max}} (|R_p|^2 + |R_s|^2) \sin\theta \, d\theta - \int_0^{\theta_{max}} (R_p^* R_s + R_p R_s^*) \sin\theta \, d\theta$$

which can be written as

$$I = I_0 \int_0^{\theta_{max}} |R_s - R_p|^2 \sin\theta \, d\theta$$

where the indices -s and -p denote the polarization of the reflected wave perpendicular or parallel, respectively, to the plane of incidence, $I_0$ is a constant independent of the optical properties of the sample, $\theta$ is the illumination angle and $\theta_{max}$ the maximal illumination angle. For isotropic multifilm structures, the coefficient of reflection can be evaluated by a 2x2 matrix method, as described by Azzam and Bashara[32] and further, Wiegand[19].

Here, As shown in Fig. 1a, a settled vesicle is modelled by a multifilm formed by 3 plane interfaces: the glass-glucose solution, the glucose solution-membrane and the membrane-sucrose solution. The variation of the intensity of the reflected light (normalized with respect to the minimal and maximal observed intensities) with the distance of the membrane to the substrate, h, is computed with this model, and is displayed in Fig. 1b and 1c for various INAs. A first remark is that the curves clearly depend on the INA even for small h values, and in particular, the value of h at the minimum light intensity increases with the INA as shown in the insert. These effects must therefore be accounted for to accurately determine the vesicle/substrate distance. An important issue shown in Fig. 1b is that at one light intensity correspond two values of h located on two distinct branches. Values on the left branch are typical of adhered membranes whereas large h-values on the right branch are characteristic of non-adhesion states. The assignment of the left value of h is easy for strongly adhered membranes which are very close to the substrate (about 10 nm). Indeed, the bright contact zone is surrounded by a dark ring observed as h increases and reaches 50 nm. However, when the membrane/substrate distance is typically of the order of 30 – 40 nm, which is the case for weak adhesion, the contact zone is dark and the contrast is not always sufficient to detect whether a darker surrounding ring exists or not. It is then very difficult to know on which branch h must be determined and, therefore, the analysis of the contact area does not allow to conclude about the adhesion state of the membrane. An attempt to overcome this problem is to consider the light intensity $I_N$ over a zone much larger than the contact zone, for instance along a radial line passing through the centre of the contact zone and to analyse the variation of $I_N$ with the radial distance over several fringes. This is enabled by the model used here, which accurately describes the damped light intensity oscillations due to the INA (Fig. 1b). The membrane/substrate distance in the central zone h, which is the only free parameter, can be then determined. This method will be used and detailed in the following.

### Experimental

### Lectin-sugar recognition

GUVs prepared with a mixture of phospholipids and lipids bearing a sugar (N-acetylglucosamine, GlcNAc) are allowed to interact with a substrate coated with lectins from Wheat Germ Agglutinin (WGA). Lectins are sugar-binding proteins which are highly specific for their sugar moieties. Lectins occur ubiquitously in nature from bacteria attachment to host cells or to specific functions in leukocyte adhesion to endothelial cells for instance through sialyl-LewisX recognition. WGA lectin is a dimer has a highly specific interaction with GlcNAc and (GlcNAc)$_2$ with two sites of recognition on each subunit. The association constant for the binding of the dimer (GlcNAc)$_2$ to WGA is rather strong[33,34], of the order of $10^4$. The affinity of the monomer GlcNAc for the WGA lectin is about 600 times weaker than that of (GlcNAc)$_2$, and is in the range of weak ligand-receptor interaction.

The adhesion strength of glyco-vesicles onto lectin-substrates strongly depends on the accessibility of the lectins grafted on the substrate to the sugar embedded in the vesicle membranes and is believed to be very small in the system presented here, probably due to geometrical considerations, so that vesicle-substrate interaction is finally ultra-weak.

**Lipids and proteins**

The lipid 1,2-Dioleoyl-sn-Glycero-3-Phosphocholine (DOPC) was purchased from Avanti Polar Lipids (Alabaster, AL, USA). The neoglycolipid was formed of Guerbet alcohol (G$_{28}$) bound to a sugar (N-acetylglucosamine) through a triethylene glycol (E$_3$) spacer (N-acetylglucosamine) according to a previously described synthesis[35] (see Fig. 2). The purity of N-acetylglucosamine derivative, GlcNAcE$_3$G$_{28}$ (M$_w$ 746.13 g/mol) was higher than 98%.

Lectin from Triticum vulgaris, synonym of wheat germ agglutinin (M$_w$ 43 000 g/mol) and FITC-labeled lectin were purchased from Sigma-Aldrich (Hannover, Germany). The heterobifunctional (sulfhydryl-selective vynilsulfone (VS) and amine selective NHS ester) polyethylene-glycol (VS – PEG –NHS, 3400 Da) was purchased from Nektar (California, USA). FITC-labeled sugar was prepared by mixing a 1μM solution of β-GlcNAc-sp-biotin (β-GlcNAc-O(CH$_2$)$_3$NHCO(CH$_2$)$_5$NH-biotin, univalent biotinylated probe) from GlycoTech (Maryland USA) with a 1μM solution of FITC-labeled avidin (Sigma). All chemicals were used without further purification.

**Vesicles**

GUVs were prepared using the standard electroformation method[36]. The lipids were dissolved in chloroform and methanol solutions (9:1 volume ratio) at 2 mg/ml. Vesicles were formed in a 200 mM sucrose solution and after storage at 5°C for a few hours, they showed quasi-spherical shapes, with marked membrane fluctuations. The vesicles were suspended in a solution (glucose and HEPES (5mM), pH=7.2) of similar osmolarity to prevent vesicle swelling and increase of membrane tension, and were observed in a cylindrical chamber (16 mm-diameter) made of a coverslip coated with grafted lectins or with silanes. Two membrane compositions were used in this study: 40% DOPC / 60% neoglycolipid (referred to as glyco-vesicles) and 100% DOPC (referred to as DOPC-vesicles).

## Substrates

Thickness corrected glass coverslides (24 mm x 24 mm, thickness 170 μm) (Assistant, Karl Hecht KG, Sondheim, Germany) were first cleaned with a detergent Decon 90 (Prolabo, France) before being immersed in a solution (Piranha) composed of 70% $H_2SO_4$ (concentration 95-98%, from Sigma) and 30% $H_2O_2$ (concentration 30%, from Sigma) for 20 minutes. They were rinsed extensively with ultra-pure (Millipore) water, dried and set in a UV – $O_3$ cleaner device for 15 minutes.

**Silanization**. The coverslides were then incubated in a solution composed of 94% acidified methanol (0.15 M acetic acid (99,5%, from Fluka, 45726 )), 4% water, and 2% (3-Mercaptopropyl) trimethoxysilane ($C_6H_{16}O_3SSi$, from Sigma M-1521) for 2 hours. They were rinsed in methanol, dried, heated at 100° C for 10 minutes, and finally rinsed in carbonate-bicarbonate buffer at pH = 8.0.

**Preparation of lectin-PEG hybrid solutions.** We mixed in equal proportion a 2 mg/ml solution (buffered at pH = 8) of lectin or FITC-lectin (36 KDa) with a 4 mg/ml solution (buffered at pH = 8) of bifunctional polyethylene-glycol polymer (VS-PEG-NHS) for 1h under gentle stirring.

**Grafting stage.** We incubated silanized coverslides in the lectin-PEG hybrid solution for 3h. Then, lectin-grafted coverslides were well rinsed 3 times in PBS buffer (pH = 7.2). All the preparation stages were performed at room temperature and stored in PBS solution at 4°C. The coverslides were used within the three days following their preparation. A schematic view of glycolipids and the substrate is presented in Fig. 2.

**Grafted lectins. Recognition by GlcNAc**

We estimated the density of grafted lectins by measuring the fluorescence of FITC-labelled grafted lectins by confocal microscopy. We found a value in the range 2000 – 2500 $\mu m^{-2}$. In order to show that GlcNAc specifically recognizes grafted lectins, we incubated grafted lectin coverslides in a solution (pH = 7.2) of glucose (200 mM) and FITC-labelled GlcNAc (1μM) for 20 min before extensive rinsing. Widefield epifluorescence images disclosed that the substrates exhibited fluorescence contrarily to bare coverslides. Comparison of the fluorescence intensity to that measured from grafted FITC-lectins revealed that about half of grafted lectins are functional for GlcNAc molecules. It is probable however that only a fraction of these functional lectins are accessible to the sugar moiety, when it is born by the lipid membrane.

**Lectin recognition of glycosites on vesicle membranes**

The specific recognition of GlcNAc on glyco-vesicle membranes by lectins in solution was tested by suspending glyco-vesicles and DOPC vesicles in a solution (200 mM glucose) containing FITC-lectins. The membranes of glyco-vesicles were fluorescent, revealing the presence of FITC-lectin fixed onto the glycolipid membrane whereas DOPC vesicles exhibited no fluorescence at their surface.

**Image acquisition and processing.**

Confocal fluorescence measurements were realized with a scanning confocal microscope Leitz, with a 1.40 NA 63x objective and blue laser excitation (488 nm). The pinhole was set to impose an optical slice thickness of about 0.4 μm.

RICM images were acquired with a Zeiss axiovert 200 inverted microscope (Carl Zeiss, Jena, Germany) equipped with a 63x antiflex objective and a C7780 camera (Hamamatsu, Tokyo, Japan). White light emitted by a HBO lamp was filtered using a green filter (λ=546+/-12 nm). 200 consecutive images (total duration 10 s, individual exposure time 50 ms) of the interference pattern were recorded for different vesicles, located at different positions on the substrate. Raw RICM images were firstly corrected for inhomogeneous illumination by a background subtraction procedure [22]. This yields an image with a uniform background as judged by comparing the intensity distribution in 20x20 pixels regions at the four corners of the image. The interference intensity of such images was considered as measured intensity (I) for further analysis.

We normalized the measured interference intensity: $I_N = (I - I_{min}) / (I_{max} - I_{min})$. Here $I_{max}$ and $I_{min}$ are the maximum and the minimum light intensity in each fringe pattern and are determined by the whole grey-scale interference pattern. Specific normalization for membrane shape reconstruction will be detailed further.

We used phase contrast microscopy (magnification x63) for direct visualization and vesicles radii determination.

Image processing and data analysis were done using the image analysis software Image-J (public domain NIH) and/or the general-purpose mathematical software Igor-pro (Wavemetrix, Portland, OR, USA), IDL (ITT Visual Information Solution, Colorado, USA) and MATLAB (The Mathworks, Inc. Massachusetts, USA), using self-written routines.

**Results**

**RICM observation of vesicle adhesion**

Typical time sequences of RICM images observed on DOPC and glyco-vesicles settled on lectin and silanized surfaces are displayed in Fig. 3. Comparison of successive images provides an indication of the temporal variation of the contact area. One observes well-defined contact zones, which are determined by the first Newtonian ring. The Newtonian fringes correspond to lines of equal height of the membrane surface above the substrate. We distinguish two behaviours.

The first behaviour is observed in the contact disc of glyco-vesicles on a lectin surface (Fig. 3A and 3B). It is characterized by an homogenous central dark zone, which is well-defined and stable with time. The central dark zone is always darker than the first fringe. A darker ring around the contact area is however not clearly and systematically observed.

In striking contrast, the contact disc of glyco-vesicles on silanized substrate (Fig. 3C) as well as that of DOPC vesicles on both silanized and lectin substrates (Fig. 3D and 3E) is non-homogeneous. The 'leopard skin' texture[30], clearly observed in Fig. 3E is characterized by a length scale of several microns. The contact zone exhibits pronounced spatial and temporal intensity fluctuations due to thermally excited membrane undulations, thus disclosing that the membranes are in the unbound phase. Moreover, contrarily to the previous case, when a dark zone is observed at the centre of the contact area, it is generally less dark than the first fringe.

In order to quantitatively describe these two behaviours, we sample the light intensity of the interference pattern of the contact zone as a function of time and lateral displacement for 17 glyco-vesicles settled on lectin substrates (radii from 11 to 41 μm), 8 DOPC vesicles settled on lectin substrates (radii from 13 to 22 μm), 7 glyco-vesicles settled on silanized substrates (radii from 14 to 30 μm) and for 4 DOPC vesicles on silanized substrates (radii from 12 to 31 μm).

**Light intensity in the centre of the contact zone.** The temporal variations of light intensity averaged over 16 pixels (pixel size: 0.1 x 0.1 μm$^2$) in the very central part of the contact zone were recorded on RICM patterns. The noise is estimated by the same quantity measured out of the interference patterns. The sampling frequency is 25 frames/s. The time variations of the normalized intensity is displayed in Fig. 4A for the five vesicles shown in Fig. 3. One clearly observes that the two glyco-vesicles settled on lectin substrates are characterized by a low normalized intensity (less than 0.2) and weak fluctuations (similar to that measured on the noise). The three other curves exhibit higher values of the light intensity and much higher fluctuations.

**Probability distribution of light intensity in the whole contact area.** We have built the probability distribution function of normalized light intensity, $P(I_N)$. The function was obtained from times series (typically 200 images, 25 fps) measured in the whole vesicle contact areas over typical areas varying from 8x8 to 32x32 pixels, depending on the size of the contact area (pixel size: 0.01 $\mu m^2$). As previously, the noise is estimated out of the interference pattern. The distribution functions obtained on the 5 vesicles shown in Fig. 3 are displayed in Fig. 4B. The two glyco-vesicles settled on lectin surfaces exhibit very similar I-distribution functions. The width of the distribution is similar to that of the noise. The distribution functions of both DOPC and glyco vesicles on silanized substrate and of DOPC vesicles on lectin substrates are shifted towards large I values and are much wider. It is worth noticing that for each vesicle, the average values of the light intensity estimated from Fig. 4A in the center of the contact area and from Fig. 4B in the whole contact area are similar: the contact area is flat on average.

The small light intensity presented by glyco-vesicles on lectin substrates compared to the other systems and the associated smaller spatial and dynamic intensity fluctuations are a strong indication for the existence of a weak homogeneous specific adhesion of glyco-vesicles onto lectin substrates. These results however concern only light intensities. In order to be conclusive, the light intensity has to be converted into a distance to the substrate. It will permit to compare the results to the values of distance reported in the literature or theoretically estimated.

**Shape reconstruction. Refined determination of membrane-substrate distance.**

We reconstruct the contour of the vesicles in the contact disc by choosing a radial line going through the centre of the disc in a direction normal to the contact line over several fringes. We recorded the light intensity on this line averaged on 10 successive images and plotted the corresponding average light intensity as a function of the radial distance. Experimental and theoretical light intensities are quantitatively compared after normalization by the same reference points. The first reference point is the maximum of light intensity, which is observed in the first bright fringe. The second reference point is the minimum light intensity, expected in the central dark zone. However, when the object is not close enough to the substrate no minimum is observed (Fig. 1). In this case, we take for the second reference point the minimum intensity reached in the second dark fringe. This intensity is unambiguously observed and, contrarily to the simplified models, is accurately computed by the model used here, which describes the damped oscillations induced by the INA.

**Spherical vesicles**. The variation of $I_N$ along a radial line is reported in Fig. 5 for two glyco-vesicles of radius equal to 18.5 $\mu m$ and 12 $\mu m$ settled on silanes and lectins respectively. The curves are fitted by the model we have described, when a spherical shape of the membrane is assumed. The fit is obtained with a single fitting parameter, the membrane – substrate distance $h_c$ at the centre of the contact zone. The two other parameters (INA and radius of the vesicle) are measured independently. As it is clearly seen in Fig. 5, the fit is very good and allows an accurate determination of $h_c$. It is equal to 90 nm and 35 nm respectively for the two displayed vesicles (glyco-vesicles on silanes and on lectins respectively).

We emphasize that this method provides a non-ambiguous measurement of $h_c$, since a whole damped sinusoidal curve is fitted with only one free parameter and we stress the importance of a careful normalization of light intensity curves, which is very sensitive to the damping induced by the numerical aperture.

**Non spherical vesicles**. When the profile of the vesicle is far from a sphere, as shown in Fig. 6, the fit of the whole $I_N(r)$ curve is no longer possible. We then calculate the theoretical curve $I_{th}(h)$ using the experimental INA. We first check that the amplitudes of the damped oscillations observed on experimental curves are similar to that on theoretical curves over at least the first three extrema of light intensity, thus indicating that experimental curves have been correctly normalized. We then assign to the membrane element located at a radial distance r characterized by the light intensity $I_N(r)$, the distance h such as $I_N(r) = I_{th}(h)$. The way we reconstructed the profile of the vesicle is illustrated in Fig. 6 on a glyco-vesicle on lectin (vesicle 3). The corresponding vesicle profile is displayed in the insert of Fig. 6. The value $h_c$ is determined from the profile extrapolated at r=0.

**Membrane fluctuations.**

The amplitude of spacing fluctuations (see Table 1) were determined from the probability distribution function of the normalized light intensity, $P(I_N)$, measured in the same way as that shown in Fig. 4B. For the non adherent cases (DOPC on lectin and silane and glyco vesicles on silane), we used the part of the curve at the right of the minimum of I(h) (Fig. 2C) to calculate membrane-substrate distances h and to derive the standard deviation of the h-distribution that we note ω. For the case of glyco-vesicles on lectin, as the fluctuations of light intensity in the contact area are of the order of magnitude of the noise (as typically shown in Fig. 4), the height fluctuations of the membrane with respect to the substrate could not be determined. They are equal or less than that due to the noise (of the order of 7 nm) and are therefore significantly smaller than that observed for the three non adherent cases.

The regime of membrane fluctuations is however dependent on the bending rigidity of the membrane, which can be slightly different for pure DOPC vesicles and glyco-vesicles. That is why we emphasize that we measured height fluctuations on the same glyco-membranes (glyco-vesicles) and that they are significantly different on lectins and on silanes. Another factor that can affect the regime of fluctuations is the membrane tension. It is directly related to to the reduced volume of the object, which depends on the vesicle preparation and is not strictly controlled. It varies over an assembly of vesicles, even when they are prepared in a single batch. That is why it is important to measure membrane fluctuations and substrate spacings on several vesicles obtained from different batches and to compare the average values obtained on each vesicle-substrate system.

**Membrane –substrate distance**

The values of the membrane-substrate distance at the centre of the contact area, $h_c$, obtained from the shape reconstructions are reported in Table 1. It clearly appears that the membrane – substrate spacing is smaller for glyco-vesicles on lectins (value of $<h_c>$ equal to 35 nm) than for glyco-vesicles on silanes (average value equal to 84 nm) and for DOPC vesicles on lectins and silanes (value of $<h_c>$ equal to 102 nm and 120 nm respectively).

The values we obtained clearly show the existence of a specific interaction between glyco-vesicles and lectin coated subtrates. Indeed, in this case, altough the average $h_c$ value is large compared to tightly adhered membranes and small compared to free membranes, it is compatible with that found on weak-adhesion patches between sialyl-Lewis[x] ligands and E-selectin receptors (30 nm)[31]. Moreover, the much higher values obtained for the three control systems can be compared to the spacing estimated for non-adherent vesicles. For a rough estimation, we write that the equilibrium membrane-substrate distance results from the balance between gravitational attraction and Helfrich repulsion[37]. Gravitation is given by $F_g = \Delta\rho\, V\, g$, where V is the vesicle volume and $\Delta\rho$ is the difference of density between the fluid in the vesicle and the suspending fluid. The Helfrich repulsion is due to the reduction of entropy induced by the presence of the substrate which reduces the amplitude of membrane thermal fluctuations. The repulsion force writes as:

$$F_{rep} = -A\frac{dV_H}{dh} = -\frac{2cA\,(kT)^2}{\kappa <h>^3}$$

where A is the contact area, $V_H$ is the Helfrich repulsive potential, c is a constant $\approx 0.115$., $\kappa$ is the membrane bending energy (we take $\kappa = 4\,10^{-20}$ J)[38], k is the Boltzmann's constant and T is the temperature. By writing that $F_g + F_{rep} = 0$ and by measuring the contact area A, we estimate the equilibrium mean spacing $<h>$. Results are presented in Table 1. We clearly see that these values are in reasonable agreement with the experimental ones for non adhering vesicles. In contrast, for glyco-vesicles on lectins, there is a factor two between predicted (70 nm) and observed (35 nm), thus bringing additional evidence to prove that the membrane-substrate distance is not determined by the gravitation/entropic repulsion balance but involves a supplementary attractive interaction.

**Discussion and conclusion**

We have worked on DOPC and glyco-vesicles prepared in a standard way (not particularly floppy). The glyco-vesicles were expected to weakly interact with lectins coated on the substrate. Observed by RICM, the contact zones of DOPC vesicles on lectin substrates and DOPC and glyco-vesicles on silanized substrates were indeed qualitatively different from that of glyco-vesicles on lectin substrates. However, in the latter case, we did not observe tight adhesion plaques, characteristic of adhered membranes. Moreover, we observed that for a wall shear-rate equal to 0.5 s$^{-1}$ the vesicles move along the substrate, indicating a very weak adhesion to the substrate. To prove the existence of this weak specific adhesion, it was necessary to use an original RICM model, which accounts for multiple incidence rays and reflection on three interfaces. It allowed us to determine the distance to the substrate and the profile of vesicles. When observations are performed at high INA, this model is particularly accurate to measure intermediate spacing in the range 30 –50 nm. The determination of these spacings for the three control systems revealed rather large values with large fluctuations, in agreement with expectations for vesicles subjected to their weight, contrarily to glyco-vesicles on lectin substrate, which were close to the substrate (35 nm in average) and displayed weak membrane fluctuations. As noted above, we see that the distance between glyco-vesicles and lectin substrate is in a range sensitive to the RICM model used for height determination. Indeed, we can compare the spacings obtained from this model to that deduced from the simplified model of normal light incidence. We found average spacings equal to 87 nm for DOPC vesicles on lectins, 102 nm and 81 nm respectively for DOPC and glyco-vesicles on silanes, which are slightly smaller than the ones obtained from the comprehensive model but do not qualitatively change the conclusion concerning membrane adhesion. For glyco-vesicles on lectins, we found an average spacing equal to 63 nm, much larger than the one found with the comprehensive model. This difference is due to several combined effects we evoked in the text: sensitivity of the $I_N(h)$ curve with the INA, uncertainty in the choice of the branch of the $I_N(h)$ curve when observations are limited to the central zone, differences of normalization between experimental and theoretical light intensities.

Although the height of glyco-vesicles on lectins derived from this simplified model is significantly smaller than that deduced on the other systems, it is however too large to conclude to membrane adhesion. The refined RICM model, which accurately describes the full $I_N(h)$ curve and allows a fit of the vesicle profile over several fringes is therefore a very suitable tool to measure membrane/substrate distances characteristic of weak adhesion states.

Finally, we select a glyco-vesicle on lectin (vesicle 1), which presents a large surface area (the radius of the contact area ≈ 3.8 mm is much larger than the, allowing the use of the method reported in reference 16 to determine its adhesion strength. We measure the contact angle and the capillary length on the vesicle profile and we find an adhesion energy per unit area of $3.2 \times 10^{-9}$ J/m$^2$, which corresponds to an adhesion energy of the order of twice the gravitational energy of the vesicle. This estimation gives an idea of the ultra-weak adhesion strength we can detect by the method we used.

In conclusion, we managed to discriminate weakly interacting glyco-vesicles from vesicles subjected to their weight. Ultra-weak interactions probably occur quite often at the vicinity of a substrate. Since they change the distance to the substrate and the fluctuations of the vesicles, it is important to detect these interactions. The method we proposed may be a way to achieve it.

**Acknowledgments.**

We thank V. Rosilio, P. Nassoy and K. Sengupta for fruitful discussions. We thank A. M. Benoliel for help for confocal experiments. Z. H. Huang gratefully acknowledges funding by the Nanolane Company. Adhesion and Inflammation and LCVN labs belong to the CNRS consortium CellTiss

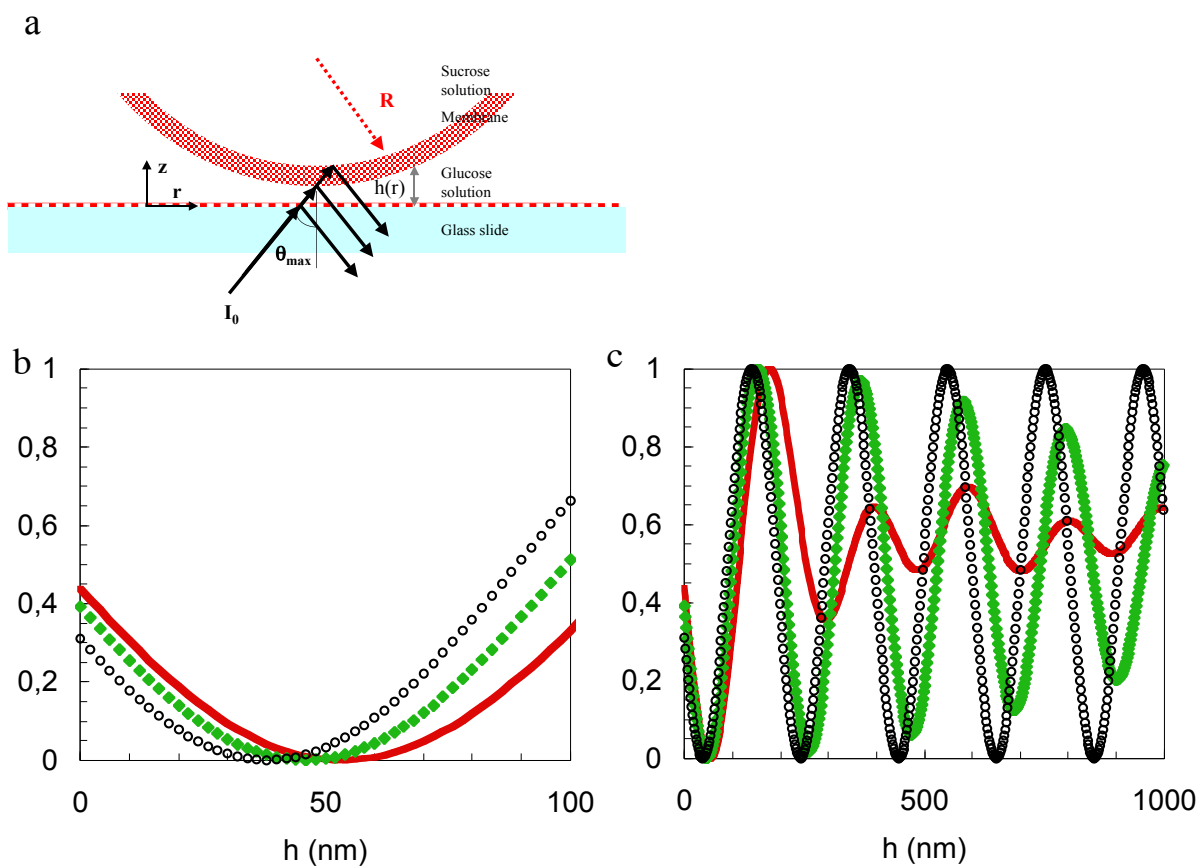

**Fig. 1. a)** Schematic representation of a vesicle at the vicinity of the substrate, and the reflections occurring at different interfaces. **b)** and **c)** Normalized relative intensity versus the substrate/vesicle distance calculated for three different values of INA (corresponding to cones of illuminating light having angles of $\theta_{max}$ ; INA = $n_{glucose}$ Sin ($\theta_{max}$) ). Black circle : normal incidence (INA= 0); Green square : INA = 0.6; Red bold line : INA = 1 ; The values of the refraction index have been taken equal to : $n_{substrate}$ = 1.525; $n_{glucose}$ = 1.3386; $n_{membrane}$ = 1.486; $n_{sucrose}$ = 1.346; The thickness of the membrane equal to 5 nm and the calcultation were done with a monochromatic wavelength of 546 nm.

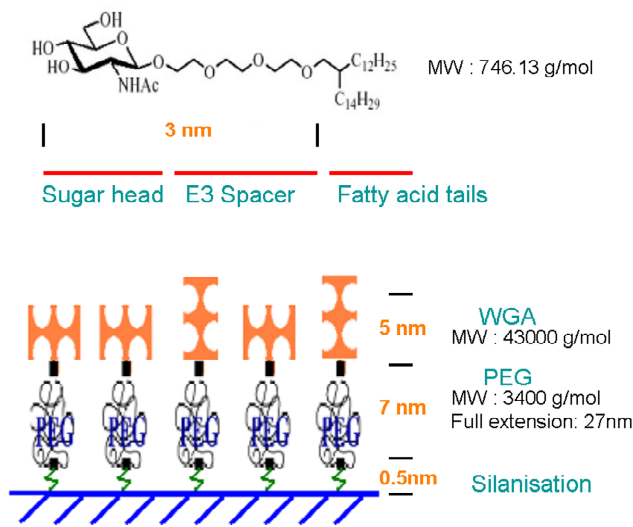

**Fig. 2.** Structure of the glycolipid and schematic drawing of the lectins grafted-molecules on the substrate

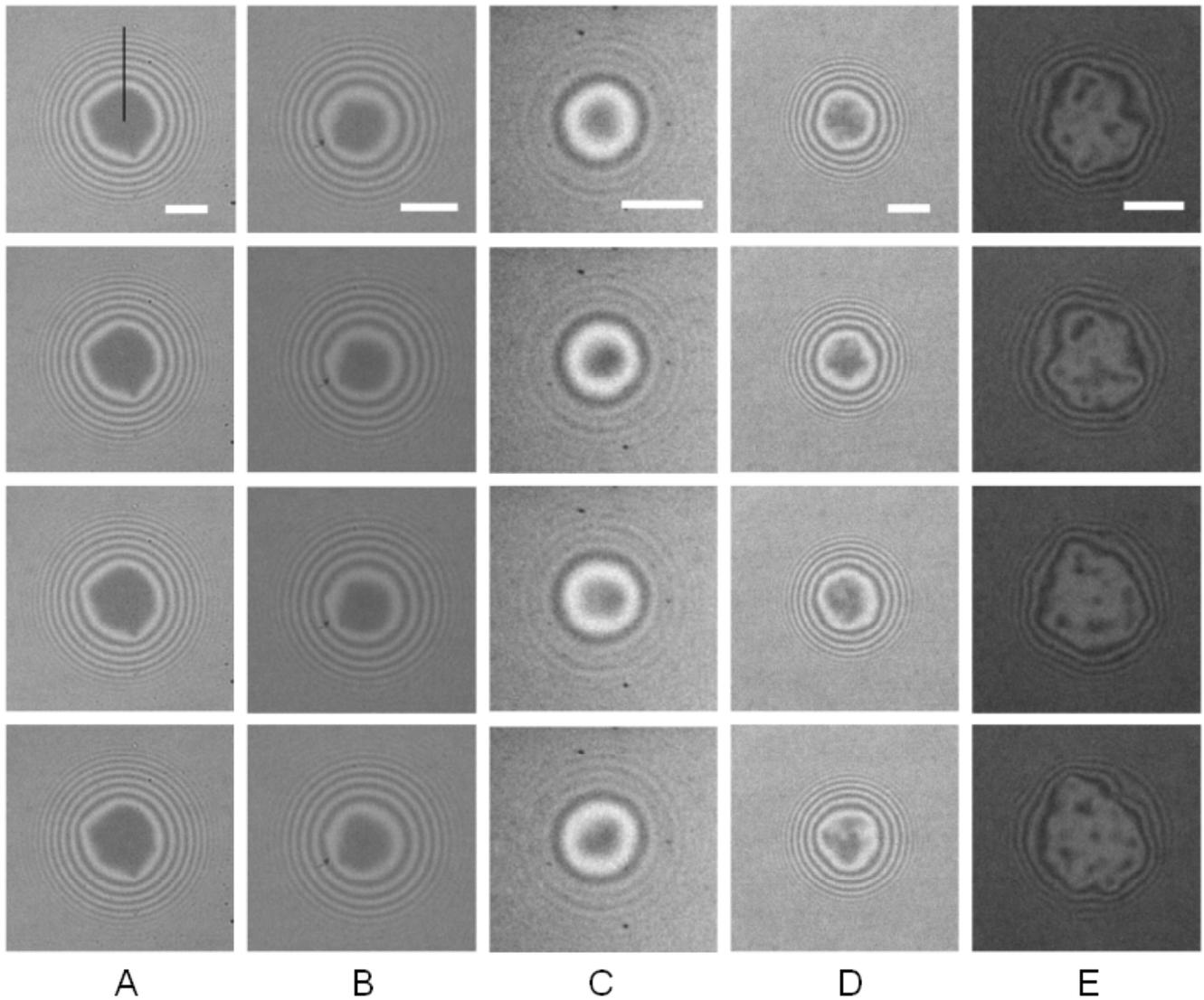

**Figure 3**. time-sequence of RICM images of 5 vesicles on different substrates: A (vesicle 1), B (vesicle 6): glyco-vesicles on lectin substrate; C (vesicle 21) : DOPC vesicle on lectin substrate; D (vesicle 27) : glyco-vesicle on silanized substrate; E (vesicle 35), DOPC vesicle on silanized substrate. Time between images was 0.2 s; white bar: 5 microns

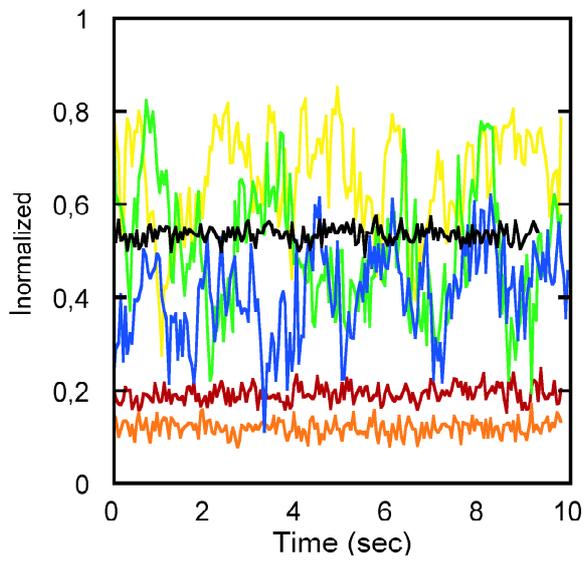
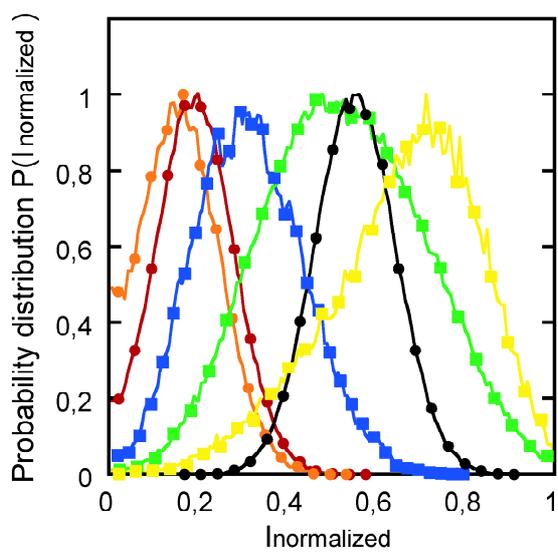

**Fig. 4**. top: time variations of the light intensity in the central part of the contact zone and, bottom: probability distribution P(I) of the light intensity over the whole contact zone for the 5 vesicles shown in Fig. 3. The noise is measured out of the zone of interference . Noise: black, vesicle A: red, vesicle B: orange, vesicle C: blue; vesicle D: green; vesicle E: yellow

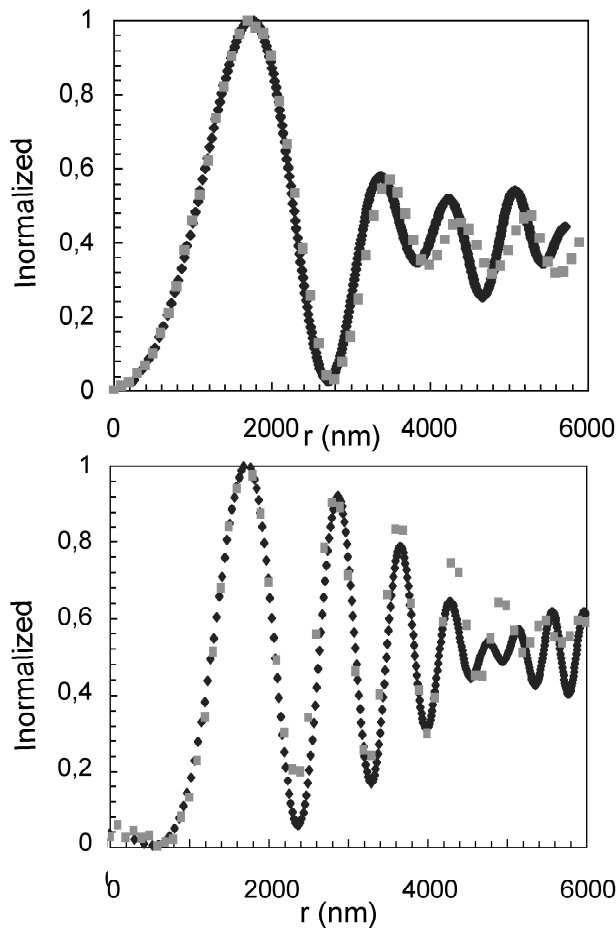

**Fig. 5 .** Variation of the normalized light intensity versus the radial distance for 2 vesicles. top :Glyco-vesicle on silanized surface (vesicle 31), R= 18.5 μm, $\theta_{max}$ = 46°, bottom : glyco-vesicle on lectin surface (vesicle 4), R = 12 μm, $\theta_{max}$ = 30° ; pink dots : experimental data, blue

curve : fit with 3 interfaces and multiple incidence model (left $h_c$ = 90 nm, right $h_c$ = 35 nm. The radial distance r writes as $r = (R^2 - (h-R + h_c)^2)^{1/2}$

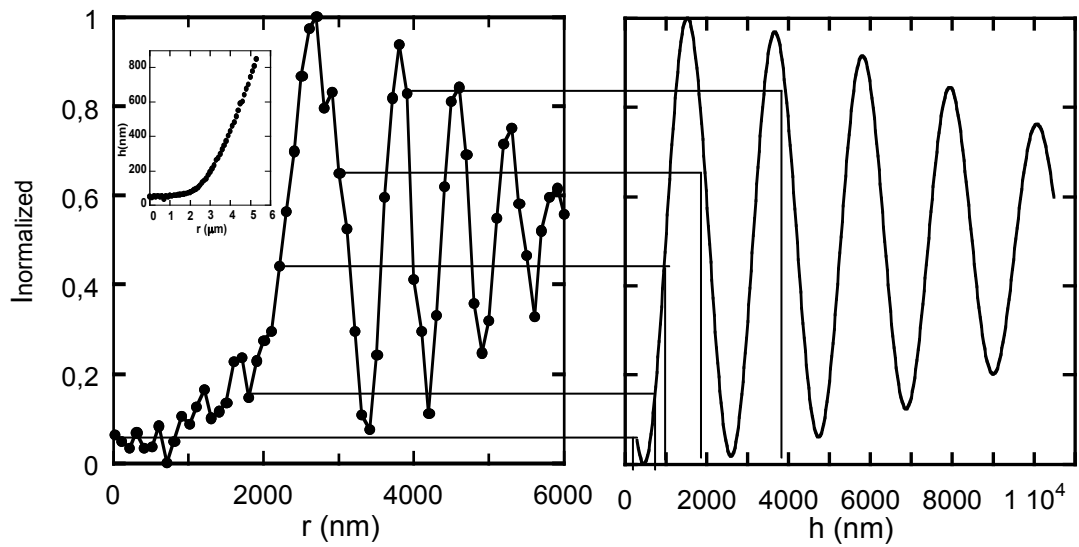

**Figure 6**. Shape reconstruction of a non spherical glyco-vesicle on lectin (vesicle 3). Left: experimental curve, right theoretical curve, stars: points used for the normalization and the intensity at r = 0. Inser: profile reconstruction

| Condition | vesicle | R (μm) | A (μm$^2$) | hc (nm) | <h> (nm) | w (nm) |
|---|---|---|---|---|---|---|
| | 1 | 18.2 | 46.4 | 15 | 110 | 6.5 |
| | 2 | 18.2 | 9.4 | 41 | 70 | 8.7 |
| | 3 | 12.1 | 19.6 | 30 | 130 | 7.4 |
| | 4 | 12.0 | 7.0 | 35 | 90 | 7.9 |
| | 5 | 14.0 | 11.8 | 30 | 90 | 7.7 |
| | 6 | 21.9 | 21.5 | 40 | 70 | 8.5 |
| | 7 | 27.2 | 24.8 | 32 | 60 | 8.1 |
| | 8 | 11.4 | 6.1 | 33 | 90 | 7.2 |
| | 9 | 25.9 | 16.7 | 47 | 60 | 7.3 |
| | 10 | 15.5 | 6.4 | 38 | 70 | 7.2 |
| | 11 | 33.3 | 23.6 | 33 | 50 | 6.8 |
| | 12 | 37.9 | 44.2 | 45 | 50 | 6.9 |
| | 13 | 11.2 | 4.0 | 39 | 80 | 8.9 |
| | 14 | 23.9 | 15.6 | 35 | 60 | 6.9 |
| glyco-vesicle on lectin coated substrate | 15 | 29.5 | 20.8 | 30 | 50 | 7.0 |
| | 16 | 14.0 | 7.5 | 35 | 80 | 9.7 |
| | 17 | 41.0 | 43.6 | 40 | 50 | 5.8 |
| Mean value | | | | 35 | 70 | 7.6 |
| Standard deviation | | | | 7.3 | 22.6 | 1.0 |
| | 18 | 19.1 | 14.9 | 113 | 70 | 8.2 |
| | 19 | 22.1 | 25.1 | 93 | 80 | 10.8 |
| | 20 | 19.1 | 11.9 | 102 | 70 | 11.1 |
| | 21 | 14.3 | 6.1 | 90 | 70 | 10.5 |
| | 22 | 13.5 | 24.8 | 98 | 120 | 13.4 |
| DOPC vesicle on lectin coated substrate | 23 | 13.0 | 11.6 | 101 | 100 | 14.1 |
| | 24 | 14.2 | 37.9 | 105 | 130 | 17.4 |
| | 25 | 19.1 | 41.5 | 115 | 100 | 12.3 |
| mean | | | | 102 | 93 | 12.2 |
| Standard deviation | | | | 8.8 | 23.8 | 2.8 |
|  | 26 | 24.1 | 12.0 | 64 | 50 | 8.7 |
|  | 27 | 16.9 | 28.9 | 94 | 100 | 12.9 |
|  | 28 | 20.7 | 24.6 | 64 | 80 | 13.1 |
|  | 29 | 29.8 | 90.9 | 84 | 90 | 12.2 |
|  | 30 | 14.8 | 4.7 | 98 | 60 | 9.4 |
| glyco-vesicle on | 31 | 18.5 | 16.6 | 90 | 50 | 11 |
| silanized substrate | 32 | 15.8 | 2.9 | 100 | 80 | 8.2 |
| Mean value |  |  |  | 84 | 73 | 10.8 |
| Standard deviation |  |  |  | 16.4 | 19.8 | 2.2 |
|  | 33 | 30.8 | 151.9 | 120 | 100 | 9.3 |
|  | 34 | 17.1 | 29.6 | 114 | 100 | 9.5 |
| DOPC vesicle on | 35 | 12.2 | 10.7 | 120 | 100 | 12.4 |
| silanized substrate | 36 | 14.0 | 25.6 | 129 | 120 | 13.7 |
| Mean value |  |  |  | 121 | 105 | 11.2 |
| Standard deviation |  |  |  | 6.2 | 10 | 2.2 |

**Table 1.** Vesicle radius, area of the contact zone, substrate-membrane spacing in the centre of the contact zone, theoretical membrane-substrate spacing calculated from equilibrium between buoyancy and Helfrich repulsion (averaged ± 5 nm), dynamic roughness for the four studied systems

70

**Notes and references**